\documentclass[aps,prl,twocolumn,preprintnumbers,amsmath,amssymb,floatfix,superscriptaddress]{revtex4-1}

\usepackage{graphics}
\usepackage{graphicx}
\usepackage{amsmath}
\usepackage{setspace}
\usepackage{epstopdf}
\usepackage{comment}
\usepackage{hyperref}
\usepackage{bm}
\usepackage{xfrac}

\hypersetup{%
   pdfpagemode=None, 
   pdfstartpage=1,
   pdfmenubar=true,
   pdftoolbar=true,
   colorlinks = true,
   linkcolor=blue,
   citecolor=blue,
   urlcolor=blue,
   bookmarksopen=false
 }

\newcommand{\affA}{Van der Waals-Zeeman Institute, Institute of Physics (IoP), University of Amsterdam, Science Park 904, 1098 XH Amsterdam, The Netherlands}
\newcommand{\affB}{Centre of New Technologies, University of Warsaw, Banacha 2c, 02-097 Warsaw, Poland}

\begin{document}

\title{Observation of collisions between cold Li atoms and Yb$^+$ ions}

\author{J.~Joger}\affiliation{\affA}
\author{H.~F\"urst}\affiliation{\affA}
\author{N.~Ewald}\affiliation{\affA}
\author{T.~Feldker}\affiliation{\affA}
\author{M.~Tomza}\affiliation{\affB}
\author{R.~Gerritsma}\affiliation{\affA}

\date{\today}

\begin{abstract}
We report on  the observation of cold collisions between $^6$Li atoms and Yb$^+$ ions. This combination of species has recently been proposed as the most suitable for reaching the quantum limit in hybrid atom-ion systems, due to its large mass ratio. For atoms and ions prepared in the $^2S_{1/2}$ ground state, the charge transfer and association rate is found to be at least~10$^{3}$ times smaller than the Langevin collision rate. These results confirm the excellent prospects of $^6$Li--Yb$^+$ for sympathetic cooling and quantum information applications. For ions prepared in the excited electronic states $^2P_{1/2}$, $^2D_{3/2}$ and $^2F_{7/2}$, we find that the reaction rate is dominated by charge transfer and does not depend on the ionic isotope nor the collision energy in the range $\sim$~1--120~mK. The low charge transfer rate for ground state collisions is corroborated by theory, but the $4f$ shell in the Yb$^+$ ion prevents an accurate prediction for the charge transfer rate of the $^2P_{1/2}$, $^2D_{3/2}$ and $^2F_{7/2}$ states. Using \textit{ab initio} methods of quantum chemistry we calculate the atom-ion interaction potentials up to energies of 30$\times 10^3$~cm$^{-1}$, and use these to give qualitative explanations of the observed rates.
\end{abstract}

\maketitle

\paragraph{Introduction}
\label{Sec_Intro}
-- Recent experiments on mixtures of ultracold atoms and ions~\cite{Zipkes:2010,Schmid:2010,Hall:2011,Rellergert:2011,Ravi:2012,Harter:2013,Haze:2015,Meir:2016} point the way to realizing novel platforms for studying quantum many-body physics~\cite{Doerk:2010,Bissbort:2013,Schurer:2016}, quantum chemistry~\cite{Hall:2011,Rellergert:2011,Ratschbacher:2012} and sympathetic cooling of trapped ions~\cite{Krych:2013,Meir:2016}. A major roadblock towards reaching the quantum regime in atom-ion interactions is formed by the time-dependent trapping field of the Paul trap, which limits attainable temperatures in hybrid atom-ion systems~\cite{Zipkes:2010,Schmid:2010,Nguyen:2012,Cetina:2012, Krych:2013,Chen:2014,Weckesser:2015,Meir:2016,Rouse:2017}. In particular, the micromotion of ions trapped in radio frequency traps may induce heating when short-range (Langevin) collisions with atoms occur. In reference~\cite{Cetina:2012}, it was shown theoretically that the lowest temperatures may be reached for the largest ion/atom mass ratios $m_\text{i}/m_\text{a}$. The atom-ion combination with the highest mass ratio, $m_\text{i}/m_\text{a}\approx 24\text{--}29$ which allows for straightforward laser cooling is Li--Yb$^+$. This combination reaches the $s$-wave limit at a collision energy of about $k_{\rm B}\cdot10\,\mu$K, which may be in reach with current Paul traps and state-of-the-art excess micromotion compensation~\cite{Cetina:2012}.

For the Li--Yb$^+$ combination to be useful in quantum information applications and for sympathetic cooling, ion loss rates should be small as compared to elastic collision rates. Such loss rates can occur in the form of charge transfer
$\text{Li}+\text{Yb}^+ \rightarrow \text{Li}^+ + \text{Yb}$ or molecule formation
$\text{Li}+\text{Yb}^+ \rightarrow \text{(LiYb)}^+$.
These elementary chemical reaction processes can be very efficiently studied in atom-ion systems, as the trapped ions allow for full control over their internal states and kinetic energy, while sophisticated detection schemes allow for studying the reaction products and their kinetic energy as well as branching ratios. Cold atom-ion reactions are experimentally investigated for the combinations Rb--Yb$^+$~\cite{Ratschbacher:2012}, Rb--Ba$^+$ and Rb--Rb$^+$~\cite{Schmid:2010}, Li--Ca$^+$~\cite{Saito:2017} and Rb--Sr$^+$~\cite{Meir:2016} for atom clouds in the ultracold regime. Furthermore, Yb--Yb$^+$~\cite{Grier:2009}, Rb--Ca$^+$~\cite{Hall:2011}, Ca--Yb$^+$~\cite{Rellergert:2011}, Ca--Ba$^+$~\cite{Sullivan:2012}, Na--Na$^+$~\cite{Goodman:2015}, Na--Ca$^+$~\cite{Smith:2014}, Rb--K$^+$ and Cs--Rb$^+$~\cite{Dutta:2017} are studied with atoms cooled in magneto-optical traps~(MOT).

In this letter, we present experimentally determined rates for chemical reactions between $^6$Li atoms and $^{171}$Yb$^+$, $^{174}$Yb$^+$ and $^{176}$Yb$^+$ ions in the mK regime. We show that the reaction rate for atoms and ions in the electronic ground state is at least $10^{3}$ times smaller than the Langevin collision rate, which should be sufficiently small for  sympathetic cooling and quantum applications. These findings are in agreement with recent calculations~\cite{Tomza:2015,DaSilva:2015}. Furthermore, we prepare the ions in excited electronic states and show that the main reaction process occurring in this situation is charge transfer. We find that for the $^2P_{1/2}$ and $^2F_{7/2}$ state in $^{174}$Yb$^+$ and $^{176}$Yb$^+$, the charge transfer rate approaches the Langevin collision rate, whereas for the metastable $^2D_{3/2}$ state, we find rates that are more than an order of magnitude smaller. We show that the charge transfer rate is almost independent of the collision energy, suggesting that charge transfer is to be associated with Langevin collisions. Our results do not depend strongly on the ionic isotope. Finally, we provide qualitative arguments for the observed rates using theoretical calculations of the excited states of the (LiYb)$^+$ system.

\paragraph{Experimental setup}
\label{Sec_Setup}

-- We overlap a cloud of $^6$Li atoms with a crystal of 2--5 Yb$^+$ ions for a certain interaction time to observe inelastic collision processes such as charge transfer or molecule formation.
While the former leads to ion loss, because $\text{Li}^+$ ions have a charge-to-mass ratio outside of the stability region of our Paul trap, the latter results in a trapped, dark $\text{(LiYb)}^+$ molecular ion. After the interaction, the atoms are released and we image the ion crystal.

\begin{figure}[htbp]
	\centering
	\includegraphics[width=0.95\columnwidth]{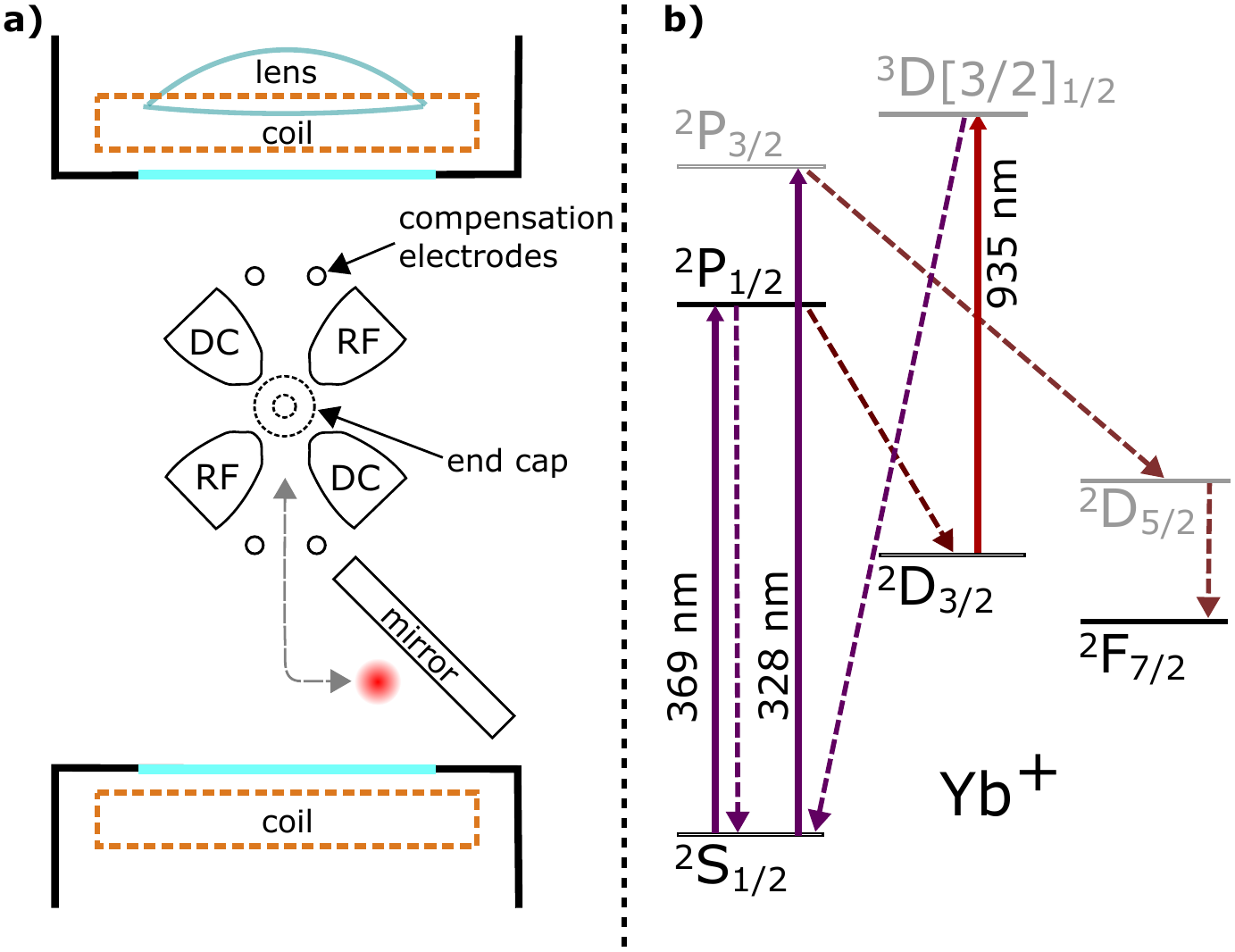}
	\caption{a) Sketch of the setup used for the combined trapping, cooling and overlapping of $^6$Li atoms and Yb$^+$ ions. The atoms are cooled in a mirror MOT about $20$\,mm below the Paul trap before they are magnetically transported into the ion trapping region. b) Level scheme of Yb$^+$. Relevant transitions and wavelengths are indicated. The states prepared for the measurement of loss rates are written in black fonts. }
	\label{fig:setup}
\end{figure}

The main part of our experimental setup, allowing to simultaneously trap and cool and to finally overlap $^6$Li and Yb$^+$ is sketched in Fig.~\ref{fig:setup} a).
Two coils in anti-Helmholtz configuration form a magnetic quadrupole field for the MOT and the magnetic trap. Dynamically adjustable currents in the two coils as well as additional compensation coils allow for a displacement of the magnetic field minimum in horizontal and vertical direction. The MOT is formed by this magnetic field and two retro-reflected circular polarized laser beams, red detuned from the $^2S_{1/2} \rightarrow \,^2P_{3/2}$ transition. One of the beams is reflected by the in-vacuum mirror under an incident angle of 45$^{\circ}$. This configuration provides confinement and cooling of the atoms in all dimensions a few cm from the Paul trap.

In 3 seconds we load about $50\times 10^6$ $^6$Li atoms from a Zeeman slower~\cite{SerwanePhD:2011} into the MOT. The atoms are cooled close to the Doppler temperature of $T_\text{D}=141\,\mu$K before they are spin polarized into the $F=3/2, m_F=3/2$ ground state by a pulse of circular polarized light resonant with the $^2S_{1/2} \rightarrow \,^2P_{1/2}$ transition and transferred into a magnetic trap. We trap $25\times 10^6$ atoms in the magnetic trap, at a temperature of $T \approx 180\,\mu$K. The initial size of the atom cloud in the magnetic trap is $\sigma_{\rm ax}\approx 1000\,\mu$m and $\sigma_{\rm vert}\approx 600\,\mu$m. During the transport into the Paul trap we compress the magnetic trap from an initial gradient of $g_z(0) = 0.44$\,T/m to a final gradient of $g_z(T_\text{tr}) = 2.8$\,T/m within the time $T_\text{tr}=120\,$ms. This reduces the size of the atom cloud to $\sigma_{\rm ax}=470(40)\,\mu$m and $\sigma_{\rm vert}=410(40)\,\mu$m leading to a higher atom density. Additionally, this reduces atom losses during the transport caused by geometrical cutoff at the compensation electrodes, and radio frequency (RF) evaporation in the immediate vicinity of the RF electrodes. Finally, $7\times 10^6$ atoms are trapped at the location of the ions, at a temperature of $T=0.6(2)$\,mK.

The Paul trap consists of four blade electrodes with a distance of 1.5\,mm to the trap center, two end caps with a spacing of 10\,mm and additional electrodes for compensation of stray electric fields. We apply an oscillating voltage with frequency $\Omega_\text{RF} = 2\pi \times 2$\,MHz and amplitude $V_0 = 75$\,V to two diagonally opposing blades and DC voltages of V$_\text{DC} \approx 15$\,V to the end caps, which yields radial and axial trap frequencies of $\omega_\text{rad}=2\pi \times 150\,$kHz and $\omega_\text{ax} = 2 \pi \times 45\,$kHz. 

We load a crystal of Yb$^+$ ions by isotope-selective two-step photoionization of neutral Yb atoms, and successive Doppler cooling on the $^2S_{1/2} \rightarrow \,^2P_{1/2}$ transition at 369\,nm. An additional beam at 935\,nm wavelength prevents the ions from being pumped to the metastable $^2D_{3/2}$ state which has a lifetime of $\tau = 52.7$\,ms\,~\cite{Yu:2000}. We detect the fluorescence light of the ions with a sCMOS camera as well as a photomultiplier tube (PMT). The ions' kinetic energy is composed of the thermal energy in the order of $T_{\rm Yb} \approx 4\,$mK after Doppler cooling, and the micromotion energy. We measure and compensate the micromotion in all three dimensions by using a set of complementary methods described in appendix~A.

\paragraph{Experimental sequence}
\label{Sec_ExperimentalSequence}

-- For the collision measurements we prepare the ions in the $^2S_{1/2}$ ground state, or in one of the excited states $^2P_{1/2}$, $^2D_{3/2}$, $^2F_{7/2}$, as follows: {\it i}) The $^2S_{1/2}$ ground state is prepared by switching off the 369\,nm Doppler cooling beam. The interaction time for the $^2S_{1/2}$ state collision measurement is  $\tau_S=200$\,ms.
{\it ii}) The $^2P_{1/2}$ state is not prepared as a pure state, but only a fraction of the population is in the $^2P_{1/2}$ state during Doppler cooling. We measure the $^2P_{1/2}$ state population fraction with the method described in Ref.~\cite{Ratschbacher:2012} to be $p(^2P_{1/2})=0.26(5)$. Here, the interaction time is $\tau_P=50$\,ms.
{\it iii}) The $^2D_{3/2}$ state is prepared by turning off the 935\,nm beam while constantly driving the $^2S_{1/2} \rightarrow \,^2P_{1/2}$ transition. The interaction time is set to $\tau_{D}=100$\,ms.
{\it iv}) To prepare the $^2F_{7/2}$ state we use a laser at 328 nm to drive the $^2S_{1/2} \rightarrow \,^2P_{3/2}$ transition, from where the ion can cascade to the $^2F_{7/2}$ state. Since this process relies on the decay of the $^2D_{5/2}$ state, which has a lifetime of 7.2\,ms~\cite{Taylor:1997}, the required preparation time is much longer than for the other states and therefore done before loading the atoms. Thus, the atoms interact with the ion already during the vertical transport and we correct the bare interaction time of 25\,ms to $\tau_{F}=39$\,ms as described in appendix~C. For the other electronic states, the ions are initially in the $^2S_{1/2}$ ground state during the transport of the atoms and only prepared in the desired state after arrival of the atoms.

Before and after the interaction we detect the ions by imaging the fluorescence light onto a sCMOS camera and determine the number of lost ions within the experimental run. The interaction time was set such that on average less than one ion was lost per experimental run. For a precise determination of the atom number, the atoms are transported back to their initial location and released from the magnetic trap. We apply a homogeneous magnetic field and after a time of flight of 2\,ms we take an absorption image with circular polarized light resonant on the $^2S_{1/2} \rightarrow \,^2P_{3/2}$ transition.

\paragraph{Results}
\label{Sec_Results}

\begin{figure}[b]
	\centering
	\includegraphics[width=0.45\textwidth]{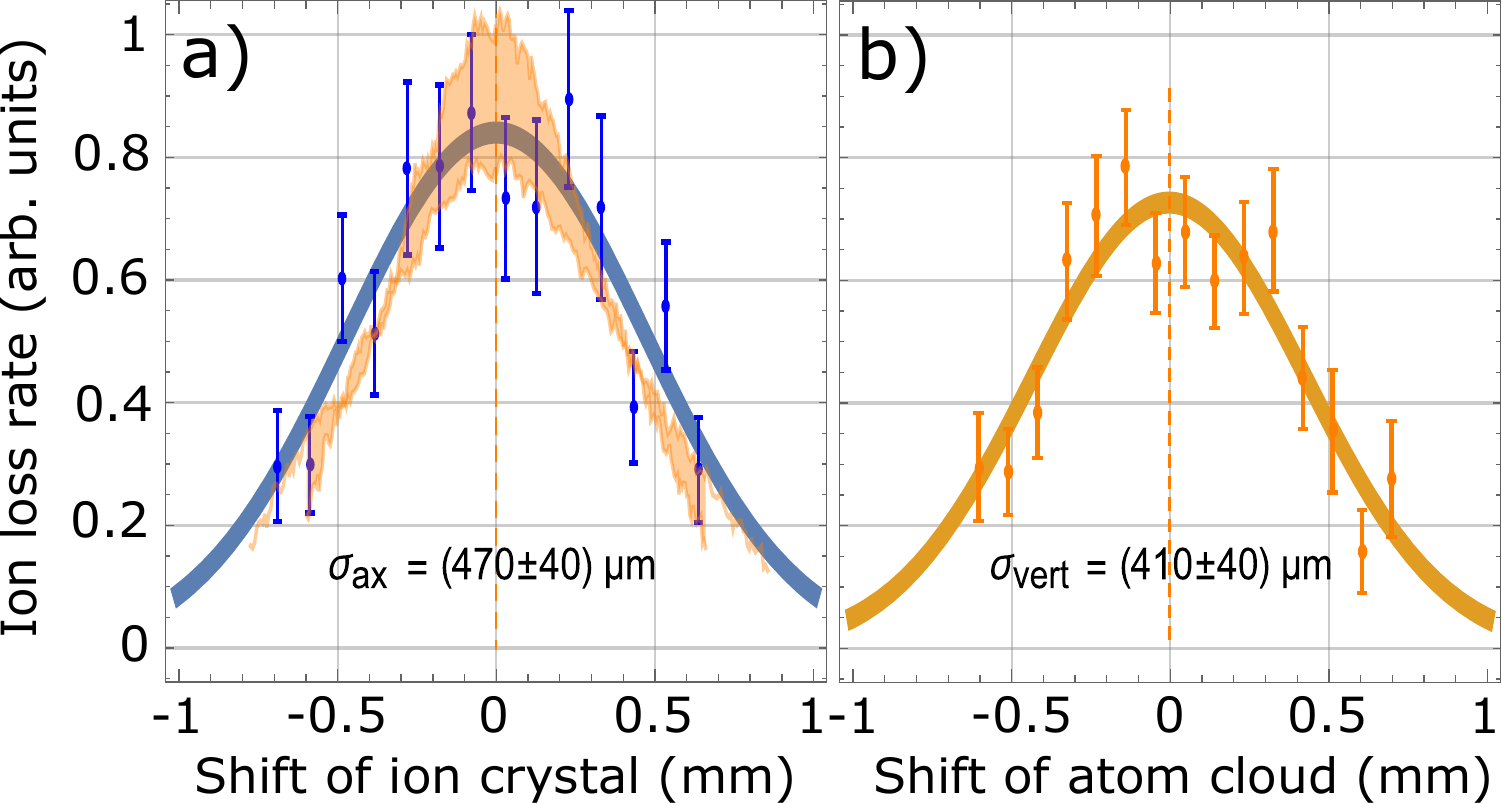}
	\caption{a) Density profile of the atom cloud along the symmetry axis of the Paul trap (magnetic field gradient $g_\text{rad} = 1.4\,$T/m). Ions are prepared in the metastable $^2D_{3/2}$ state. We shift the ions along the trap axis and measure the loss rate at each position with an interaction time of 100\,ms. The shaded region represents the atomic density profile obtained by {\it in situ} absorption imaging as explained in the text. b) Density profile in vertical direction ($g_\text{z}=2.8$\,T/m). The ions are kept in the axial center of the cloud and the atom cloud's transport elevation is scanned. Again, loss rates in the $^2D_{3/2}$ state for an interaction time of 100\,ms are measured.  }
	\label{fig:densityProfile}
\end{figure}

-- To obtain the Langevin collision rate, we first measure the spatial extent of the atom cloud by measuring ion loss rates of $^{174}$Yb$^+$ ions prepared in the metastable $^2D_{3/2}$ state as a function of the relative position within the atom cloud. The vertical profile is obtained by measuring the loss rates at different transport elevations of the magnetic trap minimum, while the horizontal profile is determined by shifting the ion string to different positions along the symmetry axis of the Paul trap by tuning end cap voltages. From the profile shown in Fig.~\ref{fig:densityProfile} we determine the size of the atom cloud to be $\sigma_\text{ax} = 470(40)\,\mu$m and $\sigma_\text{vert} = 410(40)\,\mu$m, which in combination with the atom number measurements gives us the atom density in the center of the cloud $\rho_\text{Li}=49(15)\times 10^{14}$~m$^{-3}$. From this, we determine the rate of Langevin collisions to be $\gamma_{\rm L}=2\pi \rho_\text{Li}\sqrt{C_4/\mu}=23(7)$~s$^{-1}$. Here, $C_4$ is proportional to the polarizability of the atom and $\mu$ is the reduced mass. In these measurements, the collision energy was set to $\sim$~30-50~mK. In this range, the charge transfer rate does not depend on the exact collision energy as described below. The technique to obtain the density profile is based on the assumption that the charge transfer rate is proportional to the local density of the atoms. To corroborate our measurement an {\it in situ} absorption image of the atoms at the location of the ions was made. The frequency shifts due to the magnetic trap were taken into account by combining absorption images over a frequency range of 30~MHz around the resonance. This yields $\sigma_{\text{ax}}^{in\,situ} = 378-489\,\mu$m, with the range indicating the uncertainty in the magnification of the imaging system. The atomic density obtained from this image is shown in Fig.~\ref{fig:densityProfile}(a).

From the Langevin rate and the observed ion loss rates, we obtain the fraction of Langevin collisions that lead to ion losses. The measured loss rates in units of Langevin collision rates are given in Table\,~\ref{tab:Results}. For the $^2S_{1/2}$ state only an upper bound is given as the observed loss rates are too low to be measured precisely with the currently available experimental setup.

\begin{table}[htbp]
	\centering
		\begin{tabular}{cccc}
		 \multicolumn{1}{c|}  {\textbf{~~~State~~~}} & \multicolumn{1}{c|}{\textbf{$^{174}$Yb$^+$}}& \multicolumn{1}{c|}{\textbf{$^{176}$Yb$^+$}}& \multicolumn{1}{c|}{\textbf{$^{171}$Yb$^+$}} \\[0.3ex]
			\hline
			 \multicolumn{1}{c|}{\textbf{$^2S_{1/2}$}} & \multicolumn{1}{c|}{$\leq$~10$^{-3}$} & \multicolumn{1}{c|}{$\leq$~2$\times$10$^{-4}$}& \multicolumn{1}{c|} {$\leq$~2$\times$10$^{-4}$} \\  \hline
			 \multicolumn{1}{c|}{\textbf{$^2P_{1/2}$}} & \multicolumn{1}{c|}{0.50(4)(20)}& \multicolumn{1}{c|}{0.50(6)(20)}&  \multicolumn{1}{c|}{--} \\  \hline
			 \multicolumn{1}{c|}{\textbf{$^2D_{3/2}$}} & \multicolumn{1}{c|}{0.030(1)(11)}& \multicolumn{1}{c|}{0.025(1)(10)}&  \multicolumn{1}{c|}{--} \\  \hline
			 \multicolumn{1}{c|}{\textbf{$^2F_{7/2}$}} & \multicolumn{1}{c|}{0.46(3)(16)}& \multicolumn{1}{c|}{0.41(3)(14)}&  \multicolumn{1}{c|}{--} \\  \hline		
		\end{tabular}
	\caption{Loss rates in units of Langevin collision rates for the isotopes $^{174}$Yb$^+$, $^{176}$Yb$^+$, $^{171}$Yb$^+$ in ground and excited states. For the $^2S_{1/2}$ and $^2F_{7/2}$ state, the collision energy was set to $\approx$~1 mK, whereas for the $^2P_{1/2}$ and $^2D_{3/2}$ states we took the average over the collision energy range $\approx$~1-100 mK (see Fig.~\ref{fig:crossSection}). Errors are given for statistical and systematic uncertainties. The systematic uncertainty comes from the determination of the Langevin rate $\gamma_{\rm L}$ which affects all rates equally, as well as the $^2P_{1/2}$ state population fraction for the loss rate of the $^2P_{1/2}$ state. For the ground state, only upper boundaries could be determined as the rates lie close to background ion loss. For the measurement of the $^2S_{1/2}$ state of $^{171}$Yb$^+$, the ion was prepared in the $F=0$ hyperfine ground state.}
	\label{tab:Results}
\end{table}

In principle, the methods developed in hybrid atom-ion systems also allow for studying collision-induced quenching to other electronic states~\cite{Ratschbacher:2012}. Our results did not show evidence of collision-induced quenching from the $^2F_{7/2}$ state to the electronic ground state. However, our experimental sequence does not allow for detection of quenching from the $^2P_{1/2}$ and $^2D_{3/2}$ state to the electronic ground state. During all measurements, we saw only few occurrences of dark ions and the rate is too low to distinguish these from unavoidable occurrences of $^2F_{7/2}$ states or impurity ions induced by background gas collisions. From this we conclude that charge transfer dominates over association and quenching in the $^2F_{7/2}$ state and over association in the $^2D_{3/2}$ and $^2P_{1/2}$ state.

We have also studied the energy dependence of the charge transfer rates for the $^2P_{1/2}$ and $^2D_{3/2}$ states.  The average collision energy is given by
$E_{\rm col} = \frac{\mu}{m_{\rm Yb}} \times E_{\rm Yb} +  \frac{\mu}{m_{\rm Li}} \times E_{\rm Li}$,
\noindent where $m_\text{(Li, Yb)}$ and $E_\text{(Li, Yb)}$ are the mass and kinetic energy of lithium and ytterbium, and $k_\text{B}$ the Boltzmann constant. Due to the high mass ratio $m_\text{Yb}/m_\text{Li} \approx 29$, the collision energy at low temperatures is limited by the temperature of the $^6$Li atoms ($T_\text{Li}=0.6(2)\,$mK) even though the Yb$^+$ ions' combined thermal and micromotion energy may be up to $E_{\rm Yb}/k_\text{B} \approx 10$\,mK at optimal compensation. For higher collision energies we deliberately apply a DC-electric field of up to $E_\text{DC} = 35\,$V/m in radial direction, leading to a micromotion energy of up to $E_\text{mm}/k_\text{B}=4\,$K and a collision energy of up to $E_\text{col}/k_\text{B}=120\,$mK.

\begin{figure}[t]
	\centering
	\includegraphics[width=0.45\textwidth]{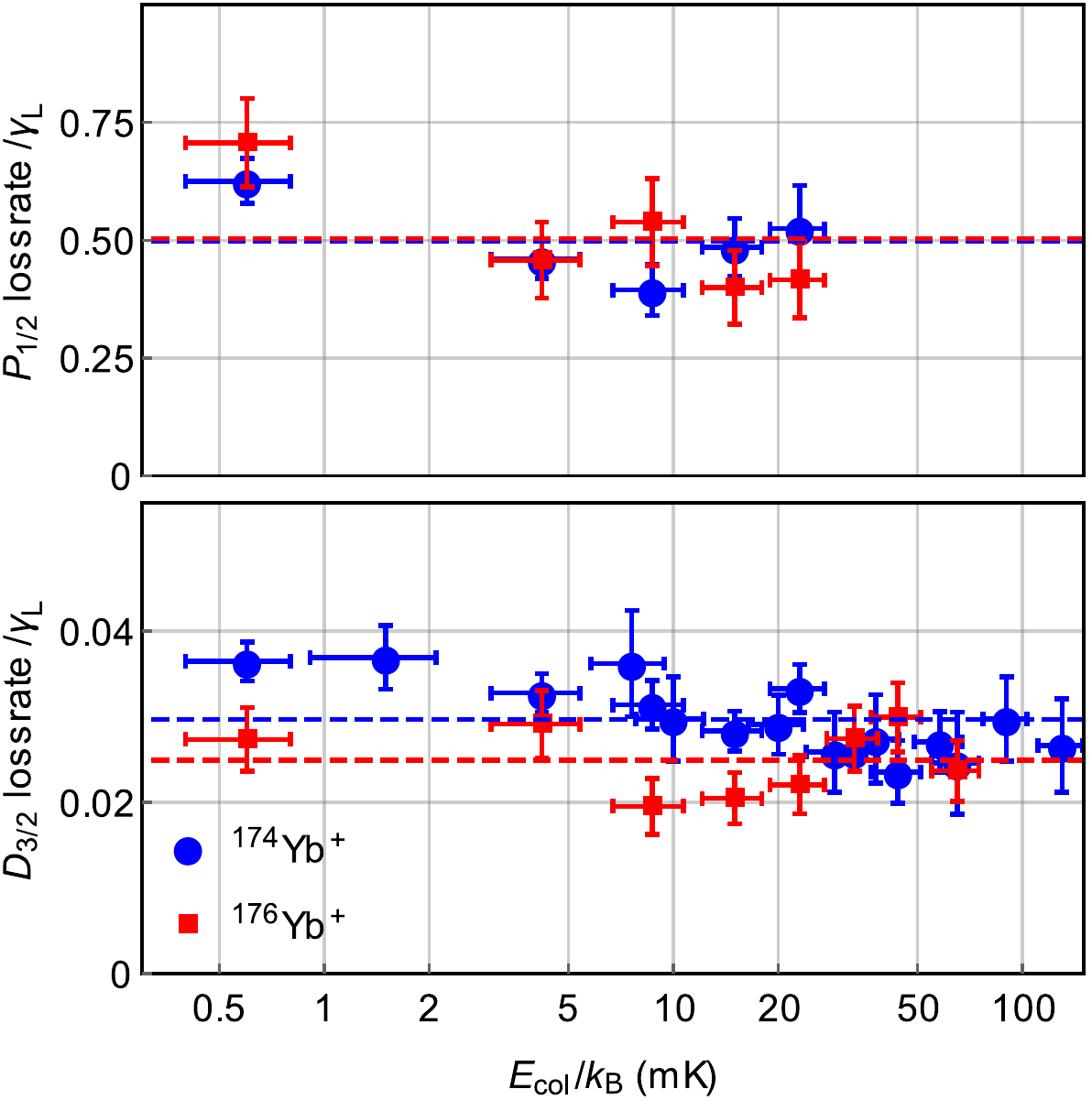}
	\caption{Collision energy dependence of ion loss rates for $^{174}$Yb$^+$ ions (blue) and $^{176}$Yb$^+$ ions (red), prepared in the $^2P_{1/2}$ state (upper) and $^2D_{3/2}$ state (lower). The dashed lines indicate the mean values of each rate.}
	\label{fig:crossSection}
\end{figure}

We measure the loss rates in the energy range of $E_{\rm col}/k_\text{B} = 1-120\,$mK. The results are shown in Fig.\,\ref{fig:crossSection}. The measured charge transfer rates are mostly independent of collision energy from which we conclude that these processes occur during Langevin collisions, whose rate is independent of collision energy~\cite{Zipkes:2010b}.

\paragraph{Theory}
\label{Sec_Theory}

-- In this section, we will present molecular structure calculations to give a qualitative understanding of the observed charge transfer rates. Supplying an \textit{ab initio} description of molecular systems containing lanthanide atoms is very challenging, because the $4f$ electrons have to be carefully included in any accurate computational model~\cite{Kotochigova:2012,Tomza:2015,Tomza:2011}. Here, we adopt and extend the computational scheme successfully applied to the ground and lowest excited states of the (LiYb)$^+$ molecular ion~\cite{Tomza:2015} and the SrYb molecule~\cite{Tomza:2011}.
The lithium atom is described by the augmented correlation consistent polarized core-valence quadruple-$\zeta$ quality basis set, aug-cc-pCVQZ~\cite{Dunning:1989}. The ytterbium atom is described using the small-core relativistic energy-consistent pseudopotential
ECP28MWB~\cite{Dolg:1989} to replace the 28 inner-shell electrons and to account for the scalar relativistic effects, and the associated basis set (15s14p12d11f 8g)/[8s8p7d7f5g] is employed~\cite{Buchachenko:2007}. Additionally, the set of $[3s3p2d1f1g]$ bond functions is used~\cite{Tao:1992}. The energy spectrum is obtained using the multireference configuration interaction method restricted to single and double exictations, MRCISD, starting from orbitals obtained with the multi-configurational self-consistent field method, MCSCF~\cite{Werner:1988}. $s$-valence orbitals and all virtual orbitals of energy up to 30$\times$10$^3\,$cm$^{-1}$ are included in the active space. Electronic structure calculations are performed with the \textsc{Molpro} package of \textit{ab initio} programs~\cite{Molpro}.

\begin{figure}[tb!]
	\begin{center}
		\includegraphics[width=\columnwidth]{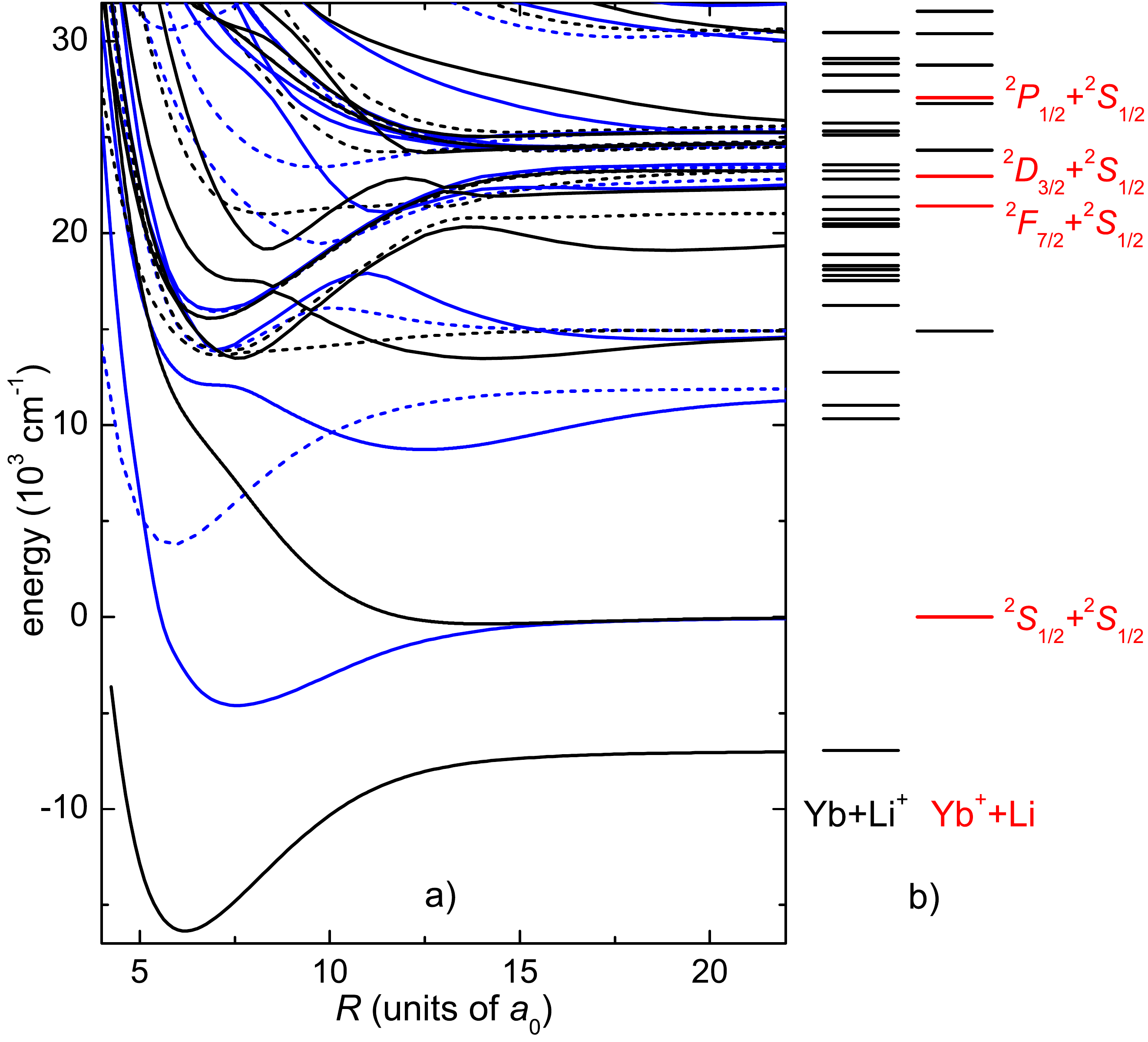}
	\end{center}
	\caption{Energy spectrum of the system: a) Nonrelativistic potential energy curves of the (LiYb)$^+$ molecular ion. Black and blue lines correspond to singlet ($S=0$) and triplet ($S=1$) states, whereas solid and dashed lines represent states with the projection of the electronic orbital angular momentum on the molecular axis equal to zero ($\Lambda=0$, $\Sigma$ symmetry) and non-zero ($|\Lambda|>0$), respectively. The lowest 5 electronic states are taken from Ref.~\cite{Tomza:2015}. b) Possible ion-atom dissociation thresholds with thresholds investigated in this work highlighted in red. Experimental energies are taken from the NIST database~\cite{NIST}.}
	\label{fig:curves}
\end{figure}

Fig.~\ref{fig:curves} presents non-relativistic electronic states of the (LiYb)$^+$ molecular ion up to an energy of about 30$\times 10^3\,$cm$^{-1}$ above the Yb$^+$($^2S_{1/2}$)+Li($^2S_{1/2}$) entrance channel. An inclusion of the spin-orbit coupling would additionally increase the number of electronic states. For comparison, Fig.~\ref{fig:curves} shows also all atomic dissociation thresholds. The large number of atomic thresholds and electronic states makes it very difficult to build a complete microscopic \textit{ab initio} model of charge transfer processes in the (LiYb)$^+$ system. Nevertheless, several conclusions can be drawn from the analysis of its electronic structure. 

The small rates of the charge transfer for Yb$^+$ ions in the $^2S_{1/2}$ state are related to the large separation of the Yb$^+$($^2S_{1/2}$)+Li(${}^2S_{1/2}$) threshold from other thresholds, reducing potential losses due to spin-orbit~\cite{Tscherbul:2016} and non-adiabatic~\cite{Sayfutyarova:2013} couplings. At the same time, the relatively small photon energy realized in radiative charge transfer reduces its Einstein coefficients.

The large rates of the charge transfer for Yb$^+$ ions in the $^2P_{1/2}$ state are due to relatively large spin-orbit coupling of the $^2P$ state ($^2P_{1/2}$--$^2P_{3/2}$ splitting over 3300$\,$cm$^{-1}$) and the large number of accessible channels for both radiative and non-radiative losses.

The relatively small charge transfer rate of the $^2D_{3/2}$ state can be related to the smaller spin-orbit coupling of the $^2D$ state, which results in a splitting between the $^2D_{3/2}$ and $^2D_{5/2}$ states of $<$~1400$\,$cm$^{-1}$. In addition, the Yb$^+$($^2D_{3/2}$)+Li($^2S_{1/2}$) threshold is surrounded by charge-transferred thresholds with a different configuration of $f$-shell electrons which should result in small coupling and mixing between related molecular electronic states.

Similarly, the large charge transfer rates for the $^2F_{7/2}$ state can be related to the very large spin-orbit coupling for the $^2F$ state which results in a splitting between the $^2F_{7/2}$ and $^2F_{5/2}$ states of $>$~10100$\,$cm$^{-1}$. Such a strong spin-orbit coupling efficiently mixes many electronic states associated with several thresholds providing opportunities for efficient radiative and non-radiative charge transfer. In contrast to the $^2D_{3/2}$ state, the Yb$^+$($^2F_{7/2}$)+Li($^2S_{1/2}$) threshold is surrounded by charge-transferred thresholds with the same configuration of $f$-shell electrons within several hundred cm$^{-1}$. Therefore, large coupling and mixing between related molecular electronic states is expected.
	
The comparison of the present results with charge transfer rates measured for Yb$^+$($^2F_{7/2}$)+Rb($^2S_{1/2}$) and Yb$^+$($^2D_{3/2}$)+Rb($^2S_{1/2}$)~\cite{Ratschbacher:2012} suggests that in fact the electronic configuration of atomic thresholds and related molecular states surrounding the entrance channel determines the short-range probability of charge transfer. In Ref.~\cite{Ratschbacher:2012}, the faster charge transfer rates were measured for Yb$^+$ ions in the $^2D_{3/2}$ state as compared to the $^2F_{7/2}$ state (opposite to the present system). However, the Yb$^+$($^2D_{3/2}$)+Rb($^2S_{1/2}$) threshold is surrounded by charge-transfered thresholds with the same configuration of $f$-shell electrons, whereas the Yb$^+$($^2F_{7/2}$)+Rb($^2S_{1/2}$) threshold is surrounded by charge-transfered thresholds with a different configuration of $f$-shell electrons (also opposite to the present system and in agreement with our qualitative explanation).

\paragraph{Conclusions}
\label{Sec_Conclusions}

-- To summarize, we have presented experimental data on cold collisions between Li atoms and Yb$^+$ ions that address two important issues:
Firstly, we have shown that reaction rates for atoms and ions prepared in their electronic ground states are at least 10$^{3}$ times smaller than the Langevin rate. This is in  agreement with theory~\cite{Tomza:2015,DaSilva:2015}.
These results are of key interest for experiments aiming to use atom-ion systems for sympathetic cooling~\cite{Krych:2010,Krych:2013,Meir:2016} and in quantum information applications~\cite{Doerk:2010,Bissbort:2013,Schurer:2016}: due to the large mass ratio Yb$^+$-Li may be the only combination that can reach the quantum regime in a Paul trap~\cite{Cetina:2012}.

Secondly, the study of cold chemical processes such as the charge transfer observed in this work, allows us to gain a deeper understanding of the molecular (Li--Yb)$^+$ system. The $4f$ shell in the Yb$^+$ ion prevents an accurate theoretical prediction of the charge transfer rate of the $^2P_{3/2}$, $^2D_{3/2}$ and $^2F_{7/2}$ states using standard methods of quantum chemistry.
Accurate measurements of chemical reactions between individual particles with full control over energy and internal states such as presented here, thus provide an excellent testbed for \textit{ab inito} molecular structure calculations. The presented results can be used to benchmark methods designed to account for the static correlation, which is related to the heavily multireference nature, and strong relativistic effects, including huge spin-orbit coupling, that are present in the exited states of the (LiYb)$^+$ molecular ion.
The development of such electronic structure methods is also relevant for systems containing other lanthanide atoms, e.g.\! erbium and dysprosium.

\begin{acknowledgements}
\section*{Acknowledgements}
This work was supported by the European Union via the European Research Council (Starting Grant 337638) and the Netherlands Organization for Scientific Research (Vidi Grant 680-47-538) (R.G.). M.T.~was supported by the National Science Centre Poland (Opus Grant~2016/23/B/ST4/03231) and PL-Grid Infrastructure. We gratefully acknowledge support from the group of Selim Jochim in Heidelberg, Germany, during the construction phase of the experimental setup.
\end{acknowledgements}


\appendix

\section{APPENDICES}

\section{A: Ion temperature determination}\label{Ap_temp}
In our paper we measure the rates of inelastic collisions between Yb$^+$ ions and $^6$Li atoms as a function of the average collision energy. This energy is given by

\begin{equation}\label{eq:colEnergy}
	E_{\rm col} = \frac{\mu}{m_{\rm Yb}} \times E_{\rm Yb} +  \frac{\mu}{m_{\rm Li}} \times E_{\rm Li} \; ,
\end{equation}

\noindent where $\mu$ is the reduced mass, and $m_\text{(Li, Yb)}$, $E_\text{(Li, Yb)}$ are the masses and kinetic energies of lithium and ytterbium, respectively. We note that the atomic densities are such that only a few Langevin collisions occur in each run and we do not expect the ionic energy to change significantly during the experiment. Thus, in order to determine the collision energy we can measure the ions' and atoms' kinetic energies separately.

\begin{figure}[b]
	\centering
	\includegraphics[width=0.45\textwidth]{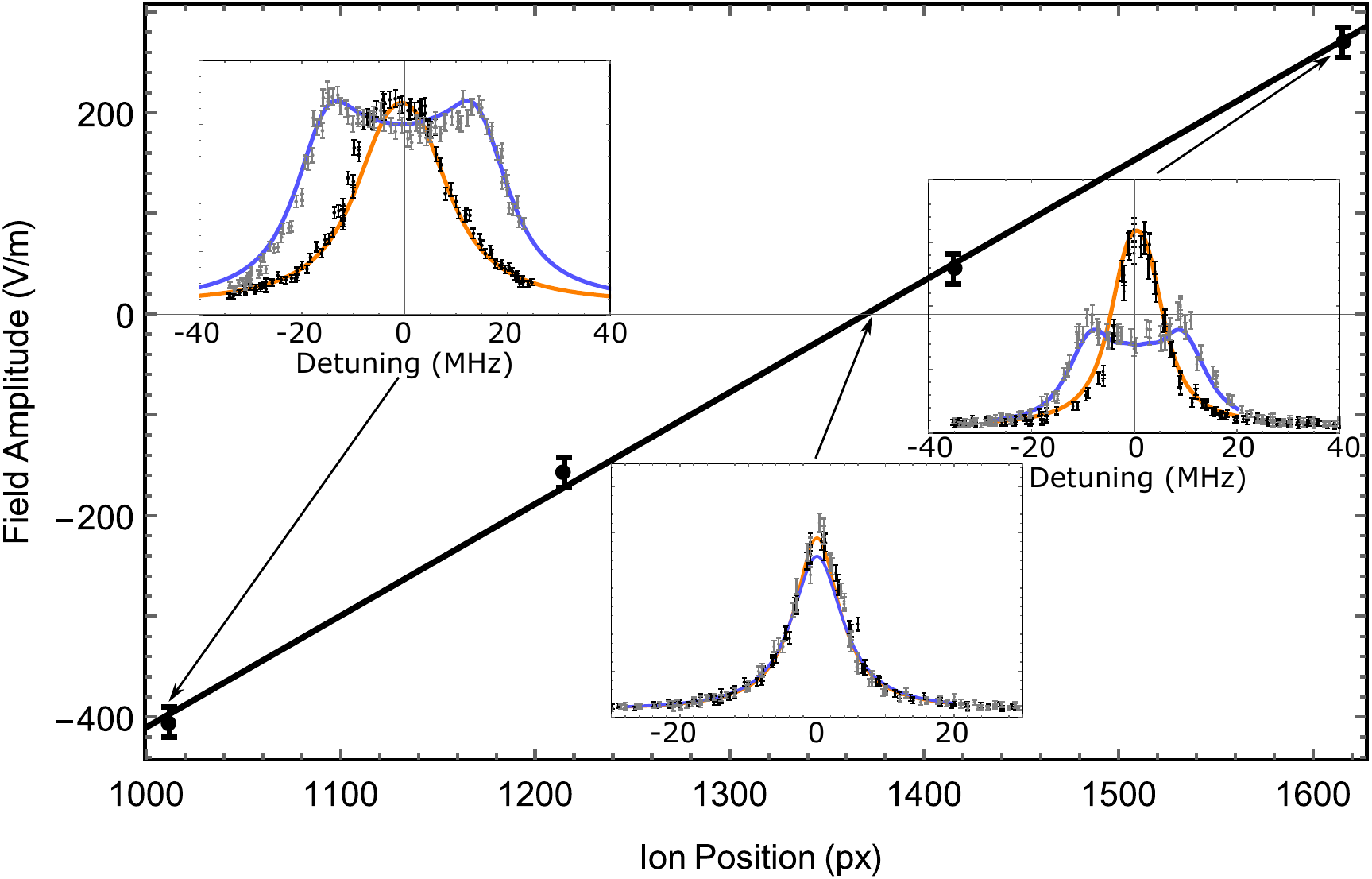}
	\caption{Oscillating electric field amplitude at 225~V drive amplitude versus axial ion position. The zero crossing corresponds to the position of minimal axial micromotion. At this position, the end cap voltages for axial confinement are 15\,V and 15.4\,V, respectively. The insets show the linewidth measurement and fit at three positions. Far away from the center, the transition is clearly broadened at 225\,V trap drive amplitude (blue) compared to the measurement at 67\,V amplitude (orange). At the center, both measurements look similar.}
	\label{fig:axMM}
\end{figure}

In a Paul trap, the motion of an ion consists of a thermal secular motion of radial and axial frequencies $\omega_{\rm rad/ax}$ in the harmonic pseudopotential, and a superimposed, driven oscillation at the trap drive frequency $\Omega_\text{RF}$ (micromotion). Thus, the ion's kinetic energy is $E_\text{Yb} = E_\text{therm}+E_\text{mm}$.
We employ microwave sideband spectroscopy on a single $^{171}$Yb$^+$ ion~\cite{Mintert:2001,Ospelkaus:2011} and infer an ion temperature of about 4~mK after Doppler cooling and a heating rate of $\leq$~4~mK/second. The excess micromotion, caused by a displacement of the ion out of the node of the dynamic electric quadrupole potential, can easily take values up to several Kelvin if not compensated carefully. Furthermore, as the micromotion amplitude may be tuned by external electric fields we can use it to control the ion's kinetic energy over a wide range.

We compensate the micromotion of the ion in all three dimensions by using a set of complementary methods~\cite{Berkeland:1998}. In axial direction, the oscillating motion of the ion is determined by the observed line broadening of an atomic transition. In both radial directions we monitor the ion's position with respect to the trap drive amplitude. From these position shifts we can deduce the DC electric field responsible for the ion shift, and thus, taking into account the trap parameters $\omega_{\rm rad}$ and $\Omega_{\rm RF}$, the micromotion amplitude.

\begin{figure}[t]
	\centering
	\includegraphics[width=0.45\textwidth]{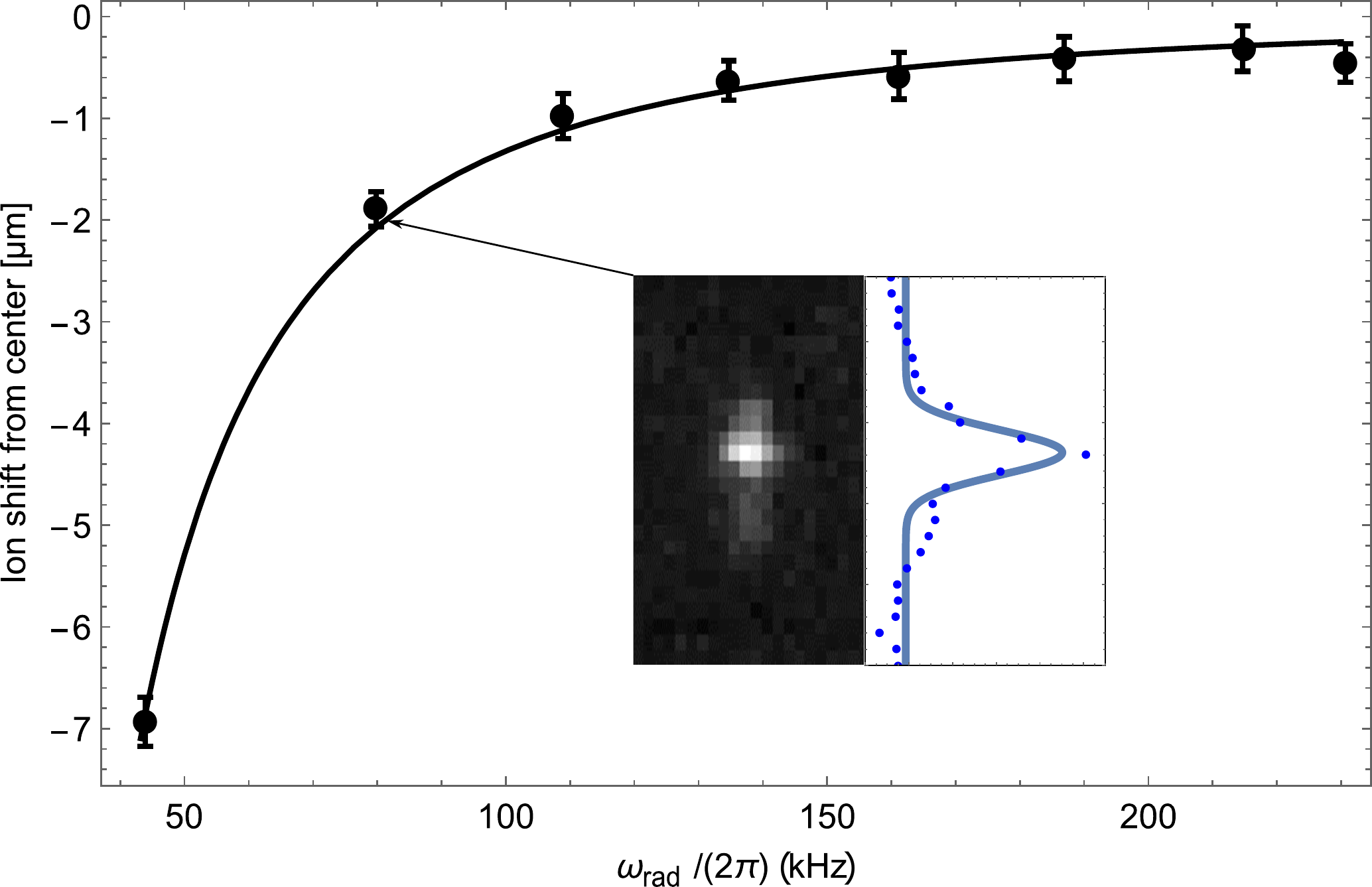}
	\caption{Radial ion position in the horizontal plane versus radial confinement. The measured positions are fitted with a model function in order to determinate the DC electric field. The insets show a camera image of an ion and a Gaussian fit to the fluorescence.}
	\label{fig:camShift}
\end{figure}

We determine the axial micromotion by measuring the line broadening of the $^2D_{3/2} \rightarrow \,^3D[3/2]_{1/2}$ transition at 935\,nm wavelength with a laser beam aligned in parallel to the trap axis. We lower the power in the laser beam in order to avoid saturation broadening. Still, the transition may be broadened by thermal motion of the ion, laser linewidth and the Zeeman splitting in a magnetic field yielding a linewidth of $\Gamma_0 > \Gamma_{\rm nat}$. Taking these effects into account, the line shape is given by~\cite{Berkeland:1998}:

\begin{equation}\label{eq:mmLineshape}
	\Gamma_{\rm mm}(\Delta) = c\times \sum J_n(\beta_{\rm mm})^2 \frac{\Gamma_0^2}{(\Delta-n\times \Omega_{\rm RF})^2 + \Gamma_0^2} \; .
\end{equation}

\noindent Here, $c$ is a constant, $J_n$ is the $n$th Bessel function, $\beta_{\rm mm} = k\times x_{\rm mm}$ is the modulation index due to the micromotion amplitude $x_{\rm mm}$ and $\Delta$ the laser detuning from resonance. For each axial position of the ion we measure the linewidth at two trap drive amplitudes of 225\,V and 67\,V. Thus, we measure the line shape at two values for $\beta_{\rm mm}$ differing by a factor of 3.3 which allows to determine $\Gamma_0$ and the micromotion-induced broadening. From a combined fit as shown in the insets of Fig.\,\ref{fig:axMM} we get a linewidth without micromotion of $\Gamma_0 \approx 2\pi \times 4\,$MHz and a position-dependent micromotion amplitude as shown in Fig.\,\ref{fig:axMM}. At a trap drive amplitude of 75\,V, as used for the loss rate measurements, we calculate an upper bound to the amplitude of the oscillating electric field in the trap center of $E_0 \leq 15\,$V$\cdot$m$^{-1}$ corresponding to a micromotion energy of $E_{\rm mm}/k_{\rm B} \leq 2.5\,$mK. At these settings, the axial field gradients can be described by an axial stability parameter $q_{\rm ax}=0.0023$, which result in axial micromotion energies of about 3~mK (0.1~mK collision energy) for the outer ions in a 5-ion crystal. Aligning the beam under 45$^{\circ}$ with respect to the trap axis allowed us to also probe for radial micromotion caused by a phase difference of the RF-signal on the two opposing electrodes, but we could not detect any.

\begin{figure}[b]
	\centering
	\includegraphics[width=0.45\textwidth]{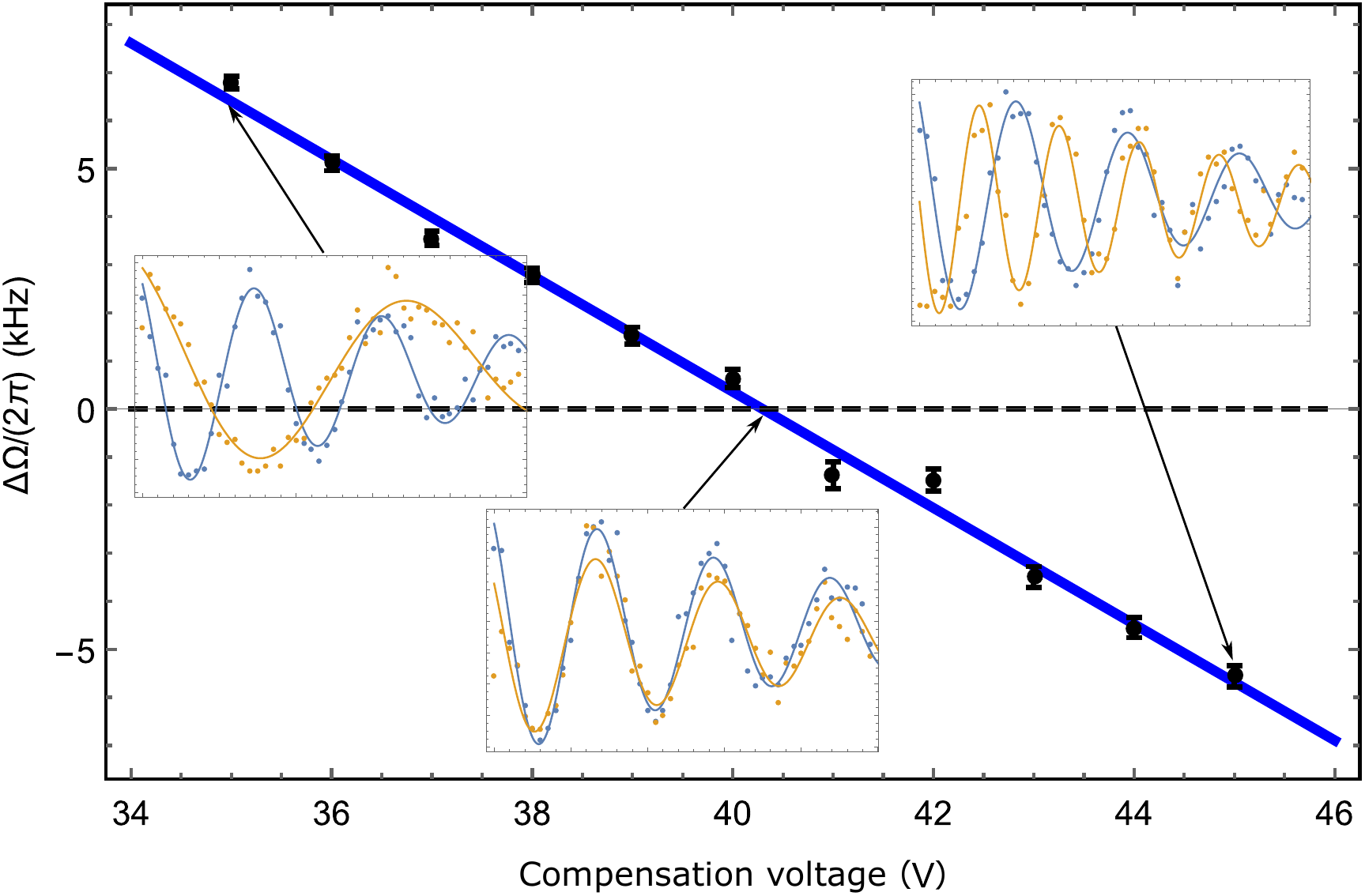}
	\caption{Radial ion position shift in the vertical plane to compensation voltage. The frequency difference of the ($^2S_{1/2},F=0,m_F=0) \leftrightarrow (^2S_{1/2},F=1,m_F=1$) transition in a magnetic field gradient is shown for different compensation voltages. The insets show the Ramsey measurements at radial trap frequencies of $\omega_\text{rad}=2\pi \times 80\,$kHz (yellow) and $\omega_\text{rad}=2\pi \times 230\,$kHz (blue). }
	\label{fig:ramseyShift}
\end{figure}

Radial excess micromotion is induced by stray electric fields which shift the ions out of the node of the RF-quadrupole field. We measure these fields by monitoring the ion's position while varying the radial confinement.
For the horizontal plane we do this by taking camera pictures while varying the radial trap frequency from $\omega_{\rm rad}=2\pi \times 40\,$kHz to $\omega_{\rm rad}=2\pi \times 220\,$kHz. We determine the ion's position with an accuracy of 200\,nm by fitting a Gaussian distribution to the camera pictures (See inset of Fig.\,\ref{fig:camShift}). By minimizing the ion's position shift we compensate the micromotion energy to about $E_{\rm mm}/k_{\rm B} \leq 2\,$mK.

In vertical direction, the ion's position cannot be determined precisely with the camera --  it is imaged from above. Instead, we use the magnetic field dependence of the ($^2S_{1/2},F=0,m_F=0) \leftrightarrow (^2S_{1/2},F=1,m_F=1$) hyperfine splitting in $^{171}$Yb$^+$ for a precise determination of position shifts. We apply a vertical magnetic field gradient of $g_{z}=0.15$\,T$\cdot$m$^{-1}$ which leads to a frequency shift of 2.1\,kHz$\cdot\mu$m$^{-1}$. Microwave Ramsey spectroscopy at radial confinements of $\omega_\text{rad}=2\pi \times 80\,$kHz and $\omega_\text{rad}=2\pi \times 230\,$kHz allows for a precise determination of the ion's position shift as shown in Fig.\,\ref{fig:ramseyShift}. From this measurement we do not only get the position of minimal micromotion, but also the micromotion energy as a function of the applied voltage at the compensation electrodes. We use this deliberate miscompensation to measure the energy dependence of the inelastic collision rates. We measure a DC electric field of $E_{\rm DC}=0.29(2)\,$V$\cdot$m$^{-1}$ for 1\,V applied to the compensation electrodes. The energy is then given by

\begin{equation}\label{key}
E_{\rm mm}(E_{\rm DC}) = \frac{E_{\rm DC}^2 \times e^2}{2\,m_\text{Yb} \times \omega_\text{rad}^2}\;,
\end{equation}

\noindent where $e$ is the elementary charge. At optimal compensation we get $E_{\rm mm}/k_\text{B} \leq 2$\,mK.

Due to the large mass ratio of $m_{\rm Yb}/m_{\rm Li} \approx 29$, the collision energy at low temperatures is almost entirely limited by the temperature of the $^6$Li atoms even though the Yb$^+$ ions' combined thermal and micromotion energy may be up to $E_{\rm Yb}/k_{\rm B} \approx 10$\,mK at optimal compensation. For higher collision energies we deliberately apply a DC electric field of up to $E_{\rm DC} = 35\,$V$\cdot$m$^{-1}$ in vertical direction, which yields a micromotion energy of up to $E_{\rm mm}/k_{\rm B}\approx 4\,$K and a collision energy of up to $E_{\rm col}/k_{\rm B}\approx 120\,$mK.

We measured the radial micromotion at various positions along the axial direction of the Paul trap over a range of 800~$\mu$K by transporting the ion by changing endcap voltages. We find that the radial micromotion compensation remains accurate over this full range within 30~V/m – corresponding to an additional collision energy of about~1~mK. We conclude that the micromotion along the axis of the Paul trap is dominated by axial oscillating field gradients.

\section{B: Compression and transport of the magnetic trap}\label{Ap_transport}
Due to limited optical access it is not possible to measure the temperature of the atom cloud inside the ion trap in the present setup. To have an estimate on the temperature, the transport of the atoms from the loading zone to the interaction zone is simulated using a cloud of $10^5$ atoms starting in a magnetic trap at a temperature of $180\,\mu$K at $g_z(0) = 0.44\,$T$\cdot$m$^{-1}$. The position of the trap minimum as well as the magnetic gradients during the transport are parameterized to follow a smooth polynomial function
\begin{align}
	\alpha(s) &= 10 s^3 - 15 s^4 + 6 s^5\,,   s \in [0,1]\label{eq:polys}\;,
\end{align}
ensuring a soft compression, acceleration and deceleration of the atoms due to the vanishing first and second order derivatives at the endpoints, $s=0$ and $s=1$.
The simulation predicts a temperature of the transported cloud in the interaction zone of $0.6\,$mK in a magnetic trap of $g_z(T_\text{tr}) = 2.8\,$T$\cdot$m$^{-1}$ after a transport time of $T_\text{tr}=120\,$ms. The adiabaticity of the transport is checked by simulating the transport back to the loading stage, coinciding with the initial temperature, which is in good agreement with our experimental observations.

\section{C: Correction of interaction time for the $^2F_{7/2}$ state}\label{Ap_tcorrect}
The preparation of the ions in the $^2F_{7/2}$ state relies on the decay of the metastable $^2D_{5/2}$ state and thus takes much longer than for the $^2P_{1/2}$ and $^2D_{3/2}$ state. Therefore, we prepare the $^2F_{7/2}$ state before the MOT loading stage.
Thus, the atoms already interact with the ions while being elevated into the interaction zone. We take this into account by calculating an effective interaction time, given by
\begin{align}
	\tau_F = \tau_{F,0} + \tau_{F,\text{corr}}\;,
\end{align}
consisting of the bare interaction time $\tau_{F,0}$ = 25\,ms, set in the experimental control software, and an additional term
\begin{align}
\tau_{F,\text{corr}} = 2 \int_{0}^{T_\text{ver}} e^{-\tfrac{\left(h-z_
	\text{min}(t)\right)^2}{2\sigma_\text{vert}^2}}\text{d} t\;,
\end{align}
consisting of the integral of the approaching atom cloud's vertical density profile being elevated by the trap minimum's trajectory $z_\text{min}(t)$ to the vertical position $h$ of the ion. The factor two accounts for the transport back. For the measured vertical cloud size of $\sigma_\text{vert}=410(40)\,\mu$m, we get $\tau_{F,\text{corr}} = 14(1)$\,ms and therefore an effective interaction time of 39(1)\,ms.



%

\end{document}